\documentclass[aps,twocolumn,prd,showpacs,nofootinbib]{revtex4}
\usepackage{amsmath}
\usepackage{graphicx}
\usepackage{dcolumn}
\usepackage{bm}
\usepackage{amssymb}
\usepackage{latexsym}

\def\be{\begin{equation}}
\def\ee{\end{equation}}
\def\ba{\begin{eqnarray}}
\def\ea{\end{eqnarray}}

\begin{document}

\title{Gravitational Wave During Slowly Evolving }

\author{Yun-Song Piao}

\affiliation{College of Physical Sciences, Graduate University of
Chinese Academy of Sciences, Beijing 100049, China}

\begin{abstract}

The ekpyrotic slow contraction or the slow expansion might be
responsible for the adiabatical production of the nearly scale
invariant curvature perturbation. However, the tensor perturbation
generated is generally strongly blue,
which implies that it is negligible on large scale. Thus it has
been still thought that the detection of primordial tensor
perturbation will rule out the relevant models. Here, we will
show a counterexample. We will illustrate that in a model of the slow
evolution, due to the rapid change of the gravitational coupling, both
the curvature perturbation and the tensor perturbation can be
nearly scale invariant.
The resulting ratio
of the tensor to scalar is in a regime which can be detected by
the coming or planned experiments. We argue that the
result is similar to that of inflation is because the background evolution given here
is actually conformally equivalent to the inflationary background.

\end{abstract}

\maketitle

The observations indicated that the primordial curvature perturbation is
nearly scale invariant. How generating it has been still a
significant issue, especially for single field. The curvature
perturbation on large scale consists of a constant mode and a mode
dependent of time \cite{Mukhanov}. When one of which is dominated
and scale invariant, the spectrum of perturbation will
be scale invariant.

The quadratic action of the curvature perturbation $\cal R$ is
generally given by \cite{GM},\cite{KYY},\cite{Vikman},\cite{KYY3},
\be S_2\sim \int d\eta d^3x {a^2 Q_{\cal R}\over c_{\cal
R}^2}\left({{\cal R}^\prime}^2-{c_{\cal R}^2}(\partial {\cal
R})^2\right), \ee where $Q_{\cal R}$ and $c_{\cal R}^2$ are
determined by the evolution of background field. $Q_{\cal R}>0$
and $c_{\cal R}^2>0$ are required to avoid the ghost and gradient
instabilities. The equation of $\cal R$ is \cite{Muk},\cite{KS},
\be u_k^{\prime\prime} +\left(c^2_{\cal R}
k^2-{z^{\prime\prime}_{\cal R}\over z_{\cal R}}\right) u_k = 0, \label{uk}\ee  after
defining $u_k \equiv z_{\cal R}{\cal R}_k$, where the prime is the
derivative for $\eta$, $z_{\cal R}\sim{a Q_{\cal R}^{1/2}/ c_{\cal R}}$.

The scale invariance of $\cal R$ requires \ba z_{\cal R}\sim {a
Q_{\cal R}^{1/2}\over c_{\cal R}} &\sim & {1\over \eta_*-\eta}\,\,
{for}\,\, {{\rm constant}}\,\,{ {\rm mode}}
\label{z2}\\
&or & (\eta_*-\eta)^2 \,\, {for}\,\,{ {\rm increasing}}\,\,{{\rm
mode}} \label{z1}\ea has to be satisfied, where initially $\eta\ll
-1$. In certain sense, both evolutions are dual \cite{Wands99}.
Noting the results will be different if $c_{\cal R}^2$ is changed
\cite{Picon},\cite{Piao0609},\cite{JM},\cite{KP},\cite{Kinney},
however, we will not involve it here, see \cite{Kinney1},\cite{JK1}
for recent discussions. In principle, both $a$ and $Q_{\cal R}$
can be changed, and together contribute the change of $z$.
However, only one among them is changed while another is hardly
changed might be the most interesting.

We have generally $Q_{\cal R}= M_P^2\epsilon$ for single field
action $P(X,\varphi)$ \cite{GM}. While the case is slightly
complicated for the most general single field
\cite{KYY},\cite{Vikman},\cite{KYY3}. However, as has been showed,
if $\epsilon<0$, we actually can manage to obtain $Q_{\cal R}\simeq
M_P^2|\epsilon|$ \cite{Piao1105}. When the scale factor is rapidly
changed while $\epsilon$ is nearly constant, the constant mode is
responsible for inflation \cite{MC}, while the increasing mode is
for the matter contraction
\cite{Wands99},\cite{FB},\cite{SS}.

However, the case can also be inverse.
When $a$ is slowly evolving, the scale invariant spectrum of
curvature perturbation can be adiabatically generated for $\epsilon\gg 1$
\cite{GKST} or $\epsilon \ll -1$ \cite{PZhou}. The evolution with
$\epsilon\gg 1$ corresponds the slow contraction, which is that of
ekpyrotic universe \cite{KOS}. While $\epsilon \ll -1$ gives the
slow expansion, which has been applied for island universe
\cite{island}. In certain sense, in Ref.\cite{PZhou} it was for
the first time observed that the slow expansion might
adiabatically induce the scale invariant spectrum of curvature
perturbation, see \cite{Piao0706} for that induced by the entropy
perturbation.

It is generally thought that for constant $|\epsilon|$, the result depends of the physics
around the exiting \cite{DH}. However, when $\epsilon$ is rapidly
changed, the thing is altered, see \cite{KS1} for the case of the
ekpyrotic slow contraction, and criticism for it \cite{LMV}. In general, during the slow contraction or the
slow expansion with rapidly changed $\epsilon$, the scale invariant adiabatical perturbation can be
naturally induced by its increasing mode \cite{Piao1012}, or its
constant mode \cite{KM},\cite{JK}. Though both cases give the
scale invariant spectrum, both pictures are distinct. In
\cite{Piao1012}, initially $|\epsilon|\gg 1$, and then is rapidly
decreasing, the slow evolution ends when $|\epsilon|\sim 1$. While
in \cite{KS1},\cite{KM},\cite{JK}, initially $|\epsilon|\lesssim
1$, and then is rapidly increasing. In \cite{JK1},
it is argued that only a finite range of the scale invariant mode
can be generated in the regime of the validity of perturbation
theory. Whether the conclusion is applicable to that induced by the increasing mode
\cite{Piao1105},\cite{Piao1012} is interesting for studying. However, even if
considering the constant mode, the larger range of the scale invariant
mode might be also obtained by a series of the phases of different
slow evolutions.

Not as the curvature perturbation, the tensor perturbation is
generally strongly blue during the slowly evolving, which implies
that it is negligible on large scale. Thus it has been still
thought that the detection of primordial tensor perturbation will
rule out the relevant models of the slow evolution. However, is it certain that the
tensor perturbation is not scale invariant ? We will revisit the
reason of having this result.

The quadratic action of the tensor perturbation $h_{ij}$ is \be
S_2\sim \int d\eta d^3x {a^2 Q_{T}\over
c_{T}^2}\left({h_{ij}^\prime}^2-{c_T^2}(\partial
{h_{ij}})^2\right), \ee 
where $Q_{T}>0$ and $c_{T}^2>0$ are required to avoid the ghost
and gradient instabilities. This action has same shape with that
of the curvature perturbation. Thus the scale invariance of
$h_{ij}$ requires \ba z_T\sim {a Q_{T}^{1/2}\over c_{T}} &\sim &
{1\over \eta_*-\eta}\,\, {for}\,\, {{\rm constant}}\,\,{ {\rm
mode}}
\label{z2}\\
&or & (\eta_*-\eta)^2 \,\, {for}\,\,{ {\rm increasing}}\,\,{{\rm
mode}} \label{h1}\ea has to be satisfied. The results will be
different if $c_T^2$ is changed, however, as for the curvature perturbation, we will not
involve this case here. In principle, both $a$ and $Q_T$ can be changed. $Q_T=M_P^2$ for $P(X,\varphi)$. We generally have not $Q_T\sim
Q_{\cal R}$, since $Q_{\cal R}\sim M_P^2|\epsilon|$ and
$|\epsilon|$ is rapidly changed. Thus it is impossible that during
the slow evolution, i.e. $a$ is hardly changed, both Eqs.(3) and
(6) are simultaneously satisfied. During the slow expansion, $z_{T}\sim a$ is hardly changed,
which implies
a strongly blue spectrum.

During the inflation, both ${\cal R}$ and $h_{ij}$ are scale
invariant. The reason is simple, since only $a$ is rapidly
changed, which just equally appears in $z_{\cal R}$ and $z_{T}$.
This is the crux
of the matter. This implies that the scale
invariance of both ${\cal R}$ and $h_{ij}$ can be obtained simultaneously, only
when the variable, which is required to be rapidly changed,
equally appears in $Q_{\cal R}$ and $Q_{T}$.
When $Q_{\cal R}\sim M_P^2|\epsilon|$ and $Q_T=M_P^2$ for $P(X,\varphi)$ are brought
into mind, it is intuitional that for hardly changed $a$ and constant $|\epsilon|\gg 1$, if \ba
M_{Peff} &\sim & {1\over \eta_*-\eta}\,\, {for}\,\, {{\rm
constant}}\,\,{ {\rm mode}}
\label{z2}\\
&or & (\eta_*-\eta)^2 \,\, {for}\,\,{ {\rm increasing}}\,\,{{\rm
mode}} \label{MP}\ea can be managed, both ${\cal R}$ and $h_{ij}$
are scale invariant might be obtained. We will discuss one possible implement in the following.

The change of $M_{Peff}$ can be obtained by a nonminimal
coupling of the scalar field to the gravity. The nonminimal
coupling has been investigated for a long history, e.g.\cite{FT}
for a review. Recently, the nonminimal
coupling inflation has been studied in \cite{Qiu},\cite{Tsu1}, and the most general
single field inflation in \cite{KYY3},\cite{Gao},\cite{Tsu}. However, here
the nonminimal coupling is introduced for a different purpose.
When the nonminimal coupling with the gravity is considered, we will study whether the result is actually as imagined. Here, we
only consider the case of (8), since only it is consistent with
the argument in Ref.\cite{Dvali}.

We will only consider the slow expansion. However, perhaps the similar
calculation is applicable to the slow contraction, in which a
bounce is required
\cite{Qiu1},\cite{Vikman1},\cite{CS},\cite{Newekp}. We begin with
the action \be {\cal L}\sim {M_{Peff}^2\over 2}R+{1\over
2}(\partial_\mu \varphi)^2-{1\over 4}\lambda \varphi^4 ,
\label{L}\ee where $M_{Peff}^2=\xi \varphi^2$ and $\xi$ is a constant to be determined. The sign before $(\partial \varphi)^2$ is reverse, however, as will be
showed, there are not the ghost instability for the perturbation.
The calculation is essentially similar to that in \cite{PZhou}.
The background evolution of the slow expansion is given in
\cite{PZhou},\cite{Piao0404}, \be a\sim {1\over (t_{*}-t)^{{1/
|\epsilon|}}},\,\,\,\,\, H= {1\over |\epsilon|(t_{*}-t)}
\label{H13}\ee for negative $\epsilon$ and $|\epsilon|\gg 1$. $\epsilon\ll -1$. Noting for $|\epsilon|\ll 1$, $a$ corresponds to
that of inflation. When the nonminimal coupling is
considered, we require this background can be still valid.
The equations of background and field are given in e.g.\cite{Qiu},
When $3H^2\varphi^2 \ll 6H{\dot \varphi}\varphi$ is neglected, the Fridmann equation
is simplified as \be H\simeq {-{1\over 2}{\dot \varphi}^2 + {1\over 4}\lambda
\varphi^4\over 3{d \over dt }M_{Peff}^2}\sim {1\over
|\epsilon|(t_{*}-t)}. \label{HH}\ee This implies \be \varphi\sim
\left(1+{\cal O}({1\over |\epsilon|})\right) {\sqrt{2/
\lambda}\over (t_*-t)^{1/|\epsilon|+1}}. \label{phi1}\ee $-{1\over
2}{\dot \varphi}^2 + {1\over 4}\lambda \varphi^4=0$ generally leads
$\varphi \sim {\sqrt{2\over \lambda}{1\over (t_*-t)}}$. The
deviation $\sim 1/|\epsilon|$ is required to accurately give
Eq.(\ref{HH}). Thus $\varphi\sim {1\over (t_*-t)}$ for
$\epsilon\ll -1$. The result is consistent with $3H^2\varphi^2 \ll
6H{\dot \varphi}\varphi$, since \be H\varphi \sim {1\over
|\epsilon|(t_*-t)^2}\ll {\dot \varphi} \sim {1\over (t_*-t)^2}
\label{ll}\ee for $\epsilon\ll -1$.

When Eqs.(\ref{H13}) and (\ref{phi1}) are considered,
$\xi$ can be determined by the equation of $\dot H$, which is
\be 2M_{Peff}^2{\dot H}-\left({d \over dt}M_{Peff}^2\right)H= {{\dot \varphi}^2
-{d^2 \over dt^2}M_{Peff}^2}, \ee where $M_{Peff}^2=\xi\varphi^2$.
The left side of this equation is 0 up to $1/|\epsilon|$ order, which requires
both terms in the right side should set off up to $1/|\epsilon|$ order. Thus we have \be \xi\simeq {1\over 6}(1+{1\over 3|\epsilon|}), \label{xi} \ee
which implies that for $|\epsilon|\gg 1$, $\xi=1/6$ is just that of
conformal coupling constant. Thus here the corresponding theory (\ref{L}) is actually nearly conformal invariant.


The equation of $\varphi$ approximately is ${\ddot
\varphi}\sim \lambda \varphi^3$. Thus the equation of the
perturbation $\delta\varphi$ is $\delta{\ddot
\varphi}-3\lambda\varphi^2\delta\varphi\sim 0$. The dominated
solution of $\delta\varphi$ is \be \delta \varphi \sim {1\over
(t_*-t)^2},\ee which is the result that $\varphi$ has weight 1 \cite{HK}.
This
perturbation can be resummed into a constant timeshift of
background, which thus is harmless.

There might be other fluids and also anisotropy. However, their
energies $\sim 1/a^n, n>0$ generally do not increase, since the
universe is still expanding. Thus they will not destroy the
background. Thus the solutions (\ref{H13}) and (\ref{phi1}) can be
stable.

When the slow expansion ends, the energy of field should be
released to reheat the universe. Here, the reheating is similar to
that after inflation. The evolution of hot big bang cosmology
begins after the reheating. However, $M^2_{Peff}\sim M_P^2$ before
the reheating should be set up. This might be done by considering
\be M^2_{Peff}\sim M_P^2/(1+{M_P^2\over \varphi^2}), \ee which
implies $M^2_{Peff}\sim \varphi^2$ for $\varphi\ll M_P$, and
$M^2_{Peff}\sim M_P^2$ for $\varphi\sim M_P$. Thus initially
$M_{Peff}$ is small, which implies the gravity is stronger, since
$G_{Newton}\sim 1/M_{Peff}^2$. However, this is consistent with
the argument in Ref.\cite{Dvali}. $\varphi$ is increased all along
during the slow expansion. Thus when the slow expansion is over, we have \be \varphi_{f}\sim
\sqrt{2\over \lambda}{1\over (t_*-t_{f})}\sim M_P. \ee
Thus $|t_f|\sim {\cal O}(|t_*|)\sim 1 /\sqrt{\lambda}M_P$. As will be
showed, $\lambda\sim 1/10^{10}$, which implies $|t_*|\sim
10^{5}t_P$.

The quadratic actions of ${\cal R}$ and $h_{ij}$ for the most
general single field, including the nonminimal coupling case, are
given in Ref.\cite{KYY3}. We will follow Ref.\cite{KYY3} for the
calculation and definition of both perturbations. Here, $Q_T$ is
given by $Q_T=M_{Peff}^2\simeq {\varphi^2\over 6}$, and $c_T^2=1$. Thus \be z_T= 0.5\,{a
Q_{T}^{1/2}\over c_{T}}\sim {a\over
2\sqrt{3\lambda}(\eta_*-\eta)}, \ee since $\eta\sim t^{1/|\epsilon|
+1}\sim t$ for $\epsilon\ll -1$. Thus the spectrum is nearly scale
invariant. The amplitude of tensor perturbation is given by
\be
{\cal P}_T= {k^3\over \pi^2}\left|{u_{kT}\over z_{T}}\right|^2\simeq {6\over \pi^2}\lambda, \ee where
$u_{Tk}\simeq {1\over \sqrt{2k}}{1\over k(\eta_*-\eta)}$,
all quantities are calculated around
$k^2\sim z_T^{\prime\prime}/z_T$, since this spectrum is induced
by the constant mode.

The calculation of the curvature perturbation is slightly
complicated, $Q_{\cal R}={\cal G}_{\cal R}c_{\cal R}^2$ and
$c_{\cal R}^2={{\cal F}_{\cal R}\over {\cal G}_{\cal R}}$. ${\cal
G}_{\cal R}$ is given by \ba {\cal G}_{\cal R} &= & {-{1\over
2}{\dot\varphi}^2-3M_{Peff}^2 H^2-6H{\dot \varphi}M_{Peff} M_{Peff,\varphi}
\over
(M_{Peff}^2 H+ {\dot\varphi}M_{Peff}M_{Peff,\varphi} )^2}M_{Peff}^4 \nonumber\\ & & + 3M_{Peff}^2 \nonumber\\
&\simeq & {1\over 6|\epsilon|}\varphi^2, \nonumber\ea where
$M_{Peff}^2=\xi \varphi^2$ and $\xi$ is given by (\ref{xi}). It is found that only that with $1/|\epsilon|$ order is left, since all terms
with $0/|\epsilon|$ is just set off. Thus ${\cal G}_{\cal R}>0$, independent of the sign
of $\epsilon$. However, if $M_{Peff}^2=M_P^2$, $M_{Peff,\varphi}=0$,
we have ${\cal G}_{\cal R}=M_P^2 \epsilon$. Thus if $\epsilon<0$,
${\cal G}_{\cal R}<0$. Here, the advantage of the rapidly changed
$M_{Peff}$ is obvious, because it is its change that alters the
sign of ${\cal G}_{\cal R}$, and leads ${\cal G}_{\cal R}>0$. In
similar reason, ${\cal F}_{\cal R}$ is given by \ba {\cal F}_{\cal
R} &= & {1\over a}\left({aM_{Peff}^4\over M_{Peff}^2 H+
{\dot\varphi}M_{Peff} M_{Peff,\varphi} }\right)^{.}-M_{Peff}^2 \nonumber\\
&\simeq & {1\over 6|\epsilon|}\varphi^2, \ea which implies
$c_{\cal R}^2={{\cal F}_{\cal
R}\over {\cal G}_{\cal R}}=1$ and $Q_{\cal R}= {1\over 6|\epsilon|}\varphi^2\sim {1\over (t_*-t)^2}\sim Q_{T}$. Thus if
we assume $Q_{\cal R}\sim M_{Peff}^2 $, we will have \be
M_{Peff}\sim {1\over (\eta_*-\eta)},\ee which is just claimed in (8).
$z_{\cal R}$ is given by \be z_{\cal R}= \sqrt{2}{a {Q}_{\cal
R}^{1/2}\over c_{\cal R}} \sim {\sqrt{2} a\over
\sqrt{3\lambda}|\epsilon|(\eta_*-\eta)}. \ee Thus similarly $\cal R$ is
nearly scale invariant, and is also induced by the constant mode.
The amplitude is \be {\cal P}_{\cal R}={k^3\over 2\pi^2} \left|{u_{k{\cal R}}\over z_{\cal R}}\right|^2
\simeq {3\over 8\pi^2}\lambda|\epsilon|. \label{pr}\ee  Thus $r$ is given by $r\equiv {{\cal
P}_T\over {\cal P}_{\cal R}}$, which is \be r= {16\over |\epsilon|}, \label{r}\ee which is only dependent of
$|\epsilon|$. It is interesting to notice that if we replace $|\epsilon|$ with $1/\epsilon$, Eq.(\ref{r}) is just that of minimal coupling inflation with
constant $\epsilon$. In principle, the larger $|\epsilon|$ gives the smaller $r$.
However, perhaps $|\epsilon|$ can not be arbitrary large, or
there will be the strong couple problem \cite{JK1}. Here, with
Refs.\cite{Gao},\cite{Tsu}, this can be discussed similarly in
detail, which is left for the future. In general, if $
|\epsilon|\sim 10^2$, Eq.(\ref{r}) implies $r\sim 0.16$. The
result is in a regime which can be detected by the coming or
planned experiments and CMB observations.
${\cal P}_{\cal R}^{1/2}\sim 10^{-5}$ requires $\lambda\sim
1/10^{10}$ for $|\epsilon|\sim 10^2$. Thus the only left and adjusted
parameter of the model is fixed, and there is not additional finetuning.

The freezed horizons of $\cal R$ and $h_{ij}$, outside which the perturbation modes freeze, are defined by
$c^2_{{\cal R},T}k^2\simeq z^{\prime\prime}_{{\cal R},T}/z_{{\cal R},T}$ 
, respectively.
The results are plotted in Fig.1. Here,
the perturbation mode naturally leaves the freezed horizons of ${\cal R}$ and $h_{ij}$, and the Hubble horizon. Thus this model can be
responsible for the emergence of scale invariant primordial
perturbations on large scale.

We reclarify the significance of the rapidly changed gravitational coupling. When the gravitational coupling is hardly changed,
i.e.in (\ref{L}) $\varphi^2$ is replaced with $M_P^2$, the similar calculation gives \be \ln{a}\sim
{1\over \lambda M_P^2(t_*-t)^2},\,\,\,\,\, H\sim {1\over \lambda M_P^2(t_*-t)^3},
\label{H15}\ee
which is consistent with (\ref{H13}) for
$M_P\sim {1\over  (t_*-t)}$. Thus the expansion is exponentially slow. It is found for this case that $\cal R$ is strongly blue. In pseudoconformal
universe, the result is similar \cite{HK}, though in a phase of slow
contraction.

In Refs.\cite{HK},\cite{Rubakov},\cite{CNT}, it is
imagined that the scale invariance of $\cal R$ is obtained by the
conversion of the perturbation of a light field $\chi$. This light
field has the conformal coupling $\varphi^2({\partial\chi})^2$.
The mechanism here is actually similar, however, for the graviton, since the coupling actually
equals 
\be {\cal L}\sim \varphi^2\left({{\dot h}_{ij}}^2-{1\over
a^2}(\partial {h_{ij}})^2\right), \ee
which naturally leads the scale invariance of tensor perturbation.
The cost is that the background evolution is altered, but luckily
is still slowly expanding. However, this cost is just interesting,
since it simultaneously leads the scale invariance of $\cal R$.

\begin{figure}[t]
\begin{center}
\includegraphics[width=7cm]{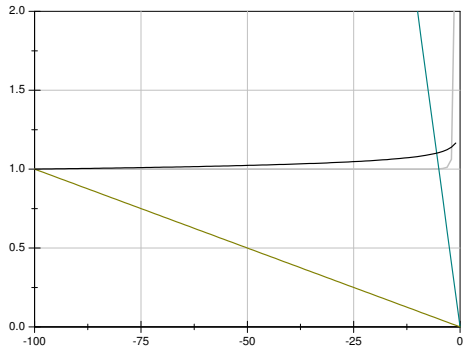}
\caption{The evolutions of $a$ for Eq.(\ref{H13}) (black line), and Eq.(\ref{H15}) (gray line) with respect to time. The freezed horizons of ${\cal R}$ and $h_{ij}$ (brown line),
and the Hubble horizon (green line) are also plotted for (\ref{H13}). We see
when the gravitational coupling is hardly changed, the expansion
is exponentially slow, while when the coupling is rapidly changed,
the exponentially slow expansion is intenerated and becomes power law. This
might be analogous with the extended inflation
\cite{Steinhardt89}.  }
\end{center}
\end{figure}

Here, the results might be incredible. However, the essential might be simple. We still
enjoy in Jordan frame, which actually can be conformally transformed to Einstein frame by the redefinition of metric. Then it can be found
the background evolution is just inflation. Thus the background given here
is actually conformally equivalent with the inflationary background.
This explains the equivalence of
result to that of inflation, since the physical predictions for observables are independent
of the frames, even if the description of backgrounds is completely different e.g.\cite{DS}.
In \cite{Gong}, it is argued that
when the universe is dominated by the single component, the perturbation is
conformal invariant fully nonperturbatively.
The conformal invariant
of perturbations implies that the different background evolution having
equivalent observable to that of inflation might be
designed by a conformal transformation of inflationary background. The model showed here is just such a
nontrivial example, which will be studied in detail in the coming.




In conclusion, it is found that, in a phase of slowly expanding, both the
curvature perturbation and the tensor perturbation can be scale
invariant, due to the rapid change of the gravitational coupling, and the resulting ratio of the tensor to scalar is in a
regime which can be detected by the coming or planned experiments.
Here, we only bring one of all possible implements of the slow
evolution, however, in principle, there could be other effective actions of
the single scalar field \cite{KYY3},\cite{DGS}, giving
the similar result,
which might be interesting for
further exploring.
This work in certain sense
highlights the fact again that identifying
the evolution of primordial universe might be a more subtle
task than expected.

\textbf{Acknowledgments} We thank Taotao Qiu for discussion. This
work is completed during KITPC program "string phenomenology and
cosmology", and supported in part by NSFC under Grant No:10775180,
11075205, in part by the Scientific Research Fund of
GUCAS(NO:055101BM03), in part by National Basic Research Program
of China, No:2010CB832804.

\end{document}